\documentclass[a4paper,aps,prl,twocolumn,floatfix,showpacs,superscriptaddress,footnoteinbib]{revtex4-1}
\usepackage{graphics}
\usepackage{epsfig}
\usepackage{times}
\usepackage{xcolor}   
\usepackage{amsfonts}
\usepackage{amsmath}
\usepackage{amssymb}
\usepackage{amsthm}
\usepackage{hyperref} 
\usepackage{wasysym}
\usepackage{bm} 
\newcommand\bloch{Bloch~}
\begin{document}
\title{\Large  Tunable multi-electron Pancharatnam phase in intensity interferometry}
\author{Disha Wadhawan}
\affiliation{Department of Physics and Astrophysics, University of Delhi, Delhi 110007}
\author{Krishanu Roychowdhury}
\affiliation{LASSP,  Department  of  Physics,  Cornell  University,  Ithaca,  NY  14853, USA}
\author{Poonam Mehta}
\affiliation{School of Physical Sciences, Jawaharlal Nehru University, New Delhi 110067 }
\author{Sourin Das}
\affiliation{Department of Physics and Astrophysics, University of Delhi, Delhi 110007}
\affiliation{Department of Physical Sciences, IISER Kolkata, Mohanpur, West Bengal 741246}
\email{dwadhawan@physics.du.ac.in, kr366@cornell.edu, pm@jnu.ac.in, sdas@physics.du.ac.in}
\date{\today}


\begin{abstract}
Pancharatnam phase was discovered in the context of polarization optics in nineteen fifties. However, its full realization in quantum many-body systems still eludes us. This is primarily due to the fact that electron spin is not easily tunable. In the present proposal, we suggest that edge states of quantum spin Hall effect (QSHE) in conjunction with spin polarized electrodes (SPE) provide us with a unique opportunity to explore the Pancharatnam phase. We demonstrate the possibility of generating and detecting the multi-electron version of this phase that can as well be interpreted as multi-particle Aharonov-Bohm (A-B) 
effect in spin space arising solely due to spin dynamics.  We further show that our proposed set-up leads to a robust interference pattern which survives orbital dephasing.
\end{abstract}
\maketitle

\noindent
{\it{\underline{Introduction}}}: 
Soon after Berry's seminal work~\cite{berry1984} which generated tremendous excitement, it was pointed out by Ramaseshan and Nityananda~\cite{rn} that the phase factor arising in cyclic changes of polarization states in Pancharatnam's work~\cite{pancharatnam1956generalized} on amplitude interferometry was in fact an early example of the Berry phase. Berry translated Pancharatnam's findings in a quantum mechanical language and introduced the Aharonov-Bohm (A-B) effect~\cite{aharonov59} on the Poincar\'e sphere by exploiting the fact that polarization of light is isomorphic to a two level quantum system~\cite{berry1987adiabatic} (see also \cite{Mehta:2009ea}). This led to wide appreciation of Pancharatnam's work in the context of geometric phases in quantum physics.  

Concurrent to this, another exciting development occurred due to Hanbury Brown and Twiss (HB-T) who replaced Michelson interferometry by intensity interferometry while measuring the diameter of stars~\cite{brown1956test}. Intensity interferometry essentially refers to processes in which a pair of particles interfere with itself. In the context of optics, a generalization of HB-T experiment was recently  proposed~\cite{mehta2010nonlocal} (a simpler set-up has been proposed recently in~\cite{chaturvedi2017classical}) which was carried out in~\cite{martin2012non,satapathy2012classical}. It was shown that 
the vector nature of light introduces a nonlocal and multi-particle geometric component  in addition to the usual dynamical component in the HB-T correlation.  
 
In the context of electronic charge transport,  nonlocal and multi-particle  AB effect has been observed in experiments involving edge currents in quantum Hall systems~\cite{neder2007interference} (see also \cite{samuelsson2004two} for theoretical developments). However it should be noted that  only the coupling to the orbital degree of freedom of electrons was exploited in \cite{neder2007interference} and the spin remained frozen.

\begin{figure}
\centering
\includegraphics[width=0.8\columnwidth]{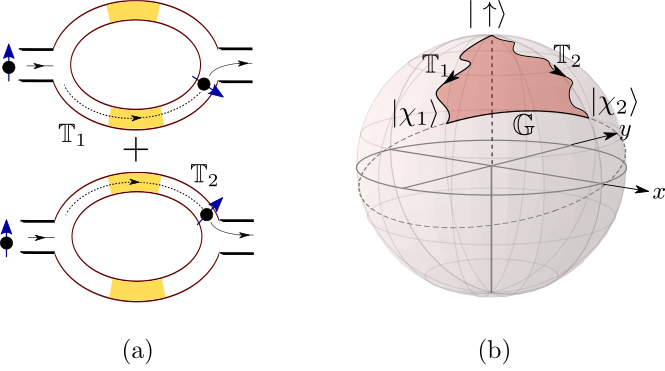}
\caption{(a) Schematic of the set-up to realize one-particle spin A-B effect. The two interfering paths are depicted as ${\mathbb{T}}_{1}$ and ${\mathbb{T}}_{2}$ and the yellow shades represent the region of rotation of spin. (b) The trajectories ${\mathbb{T}}_{1}$ and ${\mathbb{T}}_{2}$ represent the evolution of spin on the Bloch sphere. The geodesic $\mathbb{G}$ connects the end points forming a closed A-B loop surrounding the red shaded region.}
\label{fig:figg1}
\end{figure}

In the present proposal, we {{demonstrate a neat way}} to exploit the spin degree of freedom of the electrons in order to generate the A-B effect in spin space. To illustrate the idea of A-B effect in spin space, let us consider a standard two path interferometer~\cite{ji2003electronic} as a prototype. Let us further assume that the interferometer arms are endowed with the possibility of rotating the electron spin~\footnote{This can happen either due to a non-uniform magnetic field acting on the arms of the interferometer~\cite{stern1992,loss90} or due to the presence of spin-orbit coupling~\cite{datta1990electronic} in the arms of the interferometer. As the purpose of this discussion is to illustrate the essential physics of a specific kind of the A-B effect, we refrain from going into the details of the specific model responsible for inducing the rotation of the spin.} as it traverses through the respective arms [see Fig.~\ref{fig:figg1} (a)] of the interferometer. Hence, when an electron with its spin polarized along a given z-axis (call it $\vert\uparrow\rangle$) is incident on the interferometer, its amplitude of propagation will split into two parts with each part traversing coherently along the respective arm. Finally, these two amplitudes are made to  interfere producing a resulting intensity at the other end of the interferometer. Now, if we assume that the arms of the interferometer are of identical lengths with no net magnetic flux being enclosed, then one would expect a perfect constructive interference.

However, the situation changes if we allow for rotation of spin of the electron along each arm. It turns out that the spin dynamics alone can generate a nontrivial interference pattern which can be can be visualized as A-B effect on the \bloch sphere~\cite{maciejko2010spin}. Due to the spin-active interferometer arms, the incident electron with spin $\vert\uparrow\rangle$  evolves into $\vert\chi_{1}\rangle$ (lower arm) or  $\vert\chi_{2}\rangle$ (upper arm)  as it traverses  the respective arm. Hence, traversing through the lower or the upper arm actually traces out two independent trajectories [labeled ${\mathbb{T}}_1$ and ${\mathbb{T}}_2$ in Fig.~\ref{fig:figg1} (a)] starting from the same point corresponding to the incident state $\vert\uparrow\rangle$ on the \bloch sphere. Following \cite{berry1987adiabatic},  the resulting interference pattern will depend on an extra phase factor which is given by half the solid angle subtended at the centre by the closed area surrounded by ${\mathbb{T}}_1$, ${\mathbb{T}}_2$ and the geodesic~\cite{sam} ${\mathbb{G}}$ connecting $\vert\chi_{1}\rangle$ and $\vert\chi_{2}\rangle$  on this \bloch sphere. This phase is the same as the A-B  phase accumulated by an electron while traversing once around the periphery of the above defined area ($\mathcal{A}\{{\mathbb{T}}_1$,${\mathbb{T}}_2$,${\mathbb{G}}\}$) on the surface of a unit sphere [see Fig.~\ref{fig:figg1} (b)]. This can be interpreted as if a (hypothetical) monopole of strength half is sitting at the centre of this sphere~\cite{berry1984}. Hence this is referred to as an A-B effect on the \bloch sphere and the tunability of spin results in modulation of the phase which can be seen as oscillations when we change ${\mathbb{T}}_1$ or ${\mathbb{T}}_2$ or both in a controlled manner.

A set-up involving the two path interferometer type geometry which could produce such type of geometric phase from electronic spin dynamics  has been explored extensively in the past by Loss {\it et al}.~\cite{loss&goldbart} and Stern~\cite{stern1992}. In their work, the geometric phase was induced by arbitrary smooth closed loop evolution of the spin on the \bloch sphere. To this end, a question that naturally arises at the first place is, if one could as well produce this type of geometric phase in a controlled fashion resulting purely from the evolution of the electron spin only along geodesic paths on the \bloch sphere which will be a step beyond Ref.~\onlinecite{loss&goldbart, stern1992}. This will be a proper analog of Pancharatnam's geometric phase~\cite{sam} which can be visualized by considering a closed loop evolution of spins on the \bloch sphere discretized in a set of $n$ ($n > 2$) number of points on the \bloch sphere connected via geodesics hence, forming a spherical polygon. 

The questions that we address in this Letter are: (a) can we produce such a geometric phase locally in space and control it in a desired fashion without introducing an interferometer type set-up, (b) if (a) is a success, will there be any observable consequences, and finally (c), can we produce multi-electron (in our case two-electron HBT type~\cite{brown1956test}) analog (where the loop on the \bloch sphere is closed by spin evolution of not one but two electrons simultaneously) of Pancharatnam type geometric phase which is generated locally in space and measurable via standard protocols that are routinely used in electrical transport experiments in mesoscopic systems. \\


\noindent
{\it{\underline{Pancharatnam phase in amplitude interferometry} }}: First, in order to address (a) and (b), we study a set-up comprising of helical edge states (HES) of a quantum spin Hall state (QSHS) ~\cite{kane2005quantum, kane2005z, bernevig2006quantum, bernevig2006quantumnature} locally tunnel coupled to a single SPE which facilitates spin injection on the edges. Here the QSHS is hosted on the $x-y$ plane, and the spins of the helical edge states are assumed to be polarized along the z-axis with $S_z$ being conserved~\cite{brune2012spin}.

\begin{figure}
\centering
\includegraphics[width=.75\columnwidth]{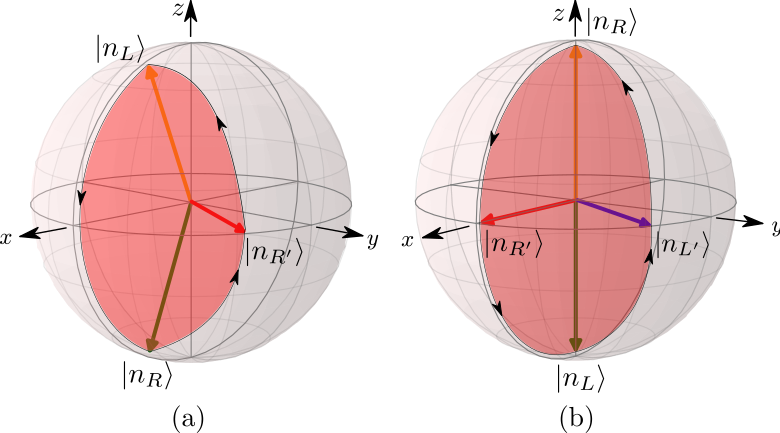}
\caption{(a) A triangular Pancharatnam loop formed by the geodesics connecting three spin states $\vert n_{R}\rangle$ , $\vert n_{L}\rangle$ and $\vert n_{R^{'}}\rangle$ on the \bloch sphere with orientation $L\rightarrow R\rightarrow R^{'}\rightarrow L$. (b) A quadrilateral Pancharatnam loop formed by the geodesics connecting four spin states $\vert n_{R}\rangle$ , $\vert n_{L}\rangle$, $\vert n_{R^{'}}\rangle$, and $\vert n_{L^{'}}\rangle$ on the \bloch sphere with orientation $R\rightarrow R'\rightarrow L\rightarrow L'\rightarrow R$. The solid angle subtended by the triangle or the quadrilateral loop at the center of the sphere is represented by $\Omega$.}
\label{fig:figg3}
\end{figure}

Dynamics of the edge states is effectively described by the Hamiltonian (assuming an intrinsic coordinate $x$ along the edge), which is valid within a linearization bandwidth, given by  $\mathcal{H}_0 = -\imath \hbar v_F \int_{-\infty}^{\infty} dx (\psi_R^\dagger \partial_x \psi_R - \psi_L^\dagger \partial_x \psi_L)$, where the operators $\psi_R^\dagger$ and $\psi_L^\dagger$ create electrons respectively for the right ($R$) and the left ($L$) propagating electron states with the spinor part of the normalized wave function given by $|n_R \rangle$ and $|n_L \rangle$ . Note that $ \langle n_L | n_R \rangle =0$ unless we break time reversal symmetry by applying an in-plane magnetic field on the QSHS~\cite{{wadhawan2016geometric}}. For simplicity, we model the SPE as a one dimensional system  whose spectrum is linearized about its Fermi energy and an unfolding trick~\cite{hou2013thermopower} is used to describe it as a right moving chiral mode ($R'$) with a specific spin polarization given by the spinor $|n_{R'} \rangle$. The corresponding Hamiltonian is $\mathcal{H}_{\rm SPE} = -\imath \hbar v_F \int_{-\infty}^{\infty} dx \psi_{R'}^\dagger \partial_x \psi_{R'}$. We further allow for weak tunneling of electrons between the SPE and the edges. 

A  finite but small backscattering within the edges is assumed to exist essentially because of possible presence of a fringing field due to proximity of ferromagnetic lead. We consider a situation where the tunneling between the SPE and the edges is local in space and it is taking place at $x=0$. Hence, the tunneling Hamiltonian is given by 
\begin{equation}
 \mathcal{H}_T = \int_{-\infty}^{\infty} dx \,  \delta(x) \big\{\sum_{\eta,\eta',\eta\neq\eta'} t_{\eta\eta'} \psi_\eta^\dagger \psi_{\eta'} + {\rm h.c.} \big\},
 \label{htun}
\end{equation}
where $\eta,\eta'\in\{R,L,R'\}$ and $t_{\eta\eta'}$ is the tunneling strength between $\eta$ and $\eta'$, further expressed as $t_{\eta\eta'}=\tilde{t}\gamma_{\eta\eta'}$ with $\gamma_{\eta\eta'}\equiv\langle n_{\eta}|n_{\eta'}\rangle$. We take the choice $\tilde{t}=t$ for $\eta,\eta'\in\{R,L\}$ ({\it i.e.} the backscattering) and $\tilde{t}=t'$ otherwise ({\it i.e.} tunneling between the SPE and the edges). Later we will consider the case of an extended tunnel junction in presence of dephasing and show that our results are robust to such consideration. \\

We now introduce the scattering matrix (or $S$-matrix) that describes the junction between the HES and the SPE. The incident wavefunction from either left contact  or right contact or the SPE on the tunnel junction at
$x=0$ is transmitted and reflected as an outgoing wavefunction. If the wavefunctions associated with the incoming and the outgoing channels are given by $\psi_{\eta}^{\rm in}$ and $\psi_{\eta}^{\rm out}$ respectively 
then the corresponding $S$-matrix elements are defined through 
\begin{equation}
 \psi^{\rm out}_\eta = \sum_{\eta'} s_{\eta\eta'}~ \psi^{\rm in}_{\eta'}.
 \label{scat_mat}
\end{equation}
We shall show below that in presence of finite backscattering ($t\neq 0$), both the current and the cross-correlated noise would feature novel oscillations arising purely from tuning the geometric phase associated with Pancharatnam loops on the \bloch sphere, which were absent for $t=0$.\\

We consider a situation where the HES of QSHS is connected to a left and right contact which are grounded {\it i.e.} $V_R=V_L=0$~($V_{L/R}$ are the voltage applied on left and right contact) while the SPE is maintained at a  bias voltage $V_{R'}= V$. In this situation the part of the total injected current into HES moving towards left or right becomes $\langle I_\eta^{\rm out} \rangle = \frac{e^2V}{h} |s_{\eta R'}|^2$, where $\eta=L/R$.  
In the weak tunneling limit between the SPE and the edge states ($t'\ll \hbar v_{F}$) we expand the current $\langle I_\eta^{\rm out} \rangle$ perturbatively up to leading order in $t'$ to obtain  
\begin{equation}
\begin{split}
\langle I_{(R/L)}^{\rm out} \rangle = \frac{e^2 V}{h} t'^2 A \{t^{2}\vert\gamma_{RL}\vert^{2}\vert\gamma_{(L/R)R^{'}}\vert^{2}+4\hbar^{2}v_{F}^{2}\vert\gamma_{(R/L)R^{'}}\vert^{2} \\
 +\, 4 \, \zeta_{(R,L)} \,t\,z \,\hbar v_{F}\, \sin(\Omega/2) \},
\end{split}
\label{Rout}
\end{equation}
in the zero temperature limit, where $A=4/(4\hbar^{2}v_{F}^{2}+t^2|\gamma_{LR}|^2)^2$, $\gamma_{RL}=\langle n_R|n_L\rangle,\gamma_{RR^{'}}=\langle n_R|n_{R^{'}}\rangle,\gamma_{R^{'}L}=\langle n_{R^{'}}|n_L\rangle$, $\zeta_R=1$, $\zeta_L=-1$, and $Z\equiv \gamma_{LR}~\gamma_{RR^{'}}~\gamma_{R^{'}L}= z e^{i\,\Omega/2}$ with $z$ being the amplitude and $\Omega/2$ being the phase of the complex number $Z$ which is the quantity of central focus. It essentially represents a series of cyclic  projections $L\rightarrow R\rightarrow R^{'}\rightarrow L$ forming a spherical triangle connected by three geodesics on the \bloch sphere [Fig.~\ref{fig:figg3} (b)]. The quantity $\Omega$ represents the solid angle subtended by this triangle at the center of the \bloch sphere and can be identified with Pancharatnam's geometric phase~\cite{sam}. It should be noted that this phase can be tuned by altering the magnetization direction of the SPE leading to coherent oscillations in the current.

For weak in-plane magnetic field applied on the QSHS such that a gap opens up on the edges while keeping the bulk intact and then by doping on the edge states, $\vert \gamma_{RL}\vert$ can be tuned to non-zero values (see Ref.~[\onlinecite{wadhawan2016geometric}]). Thus, oscillations can be induced simply by stretching the triangular Pancharatnam loops area by tuning magnetization direction of the tip alone [see Fig.~\ref{fig:figg3} (a)]. Similarly, the cross-correlated noise between the left and the right contact under the same condition as mentioned above can be obtained as~\cite{blanter2000shot}
\begin{equation}
\begin{split}
 &S_{RL} = \frac{e^3V}{\pi \hbar} t'^4 A^2 \bigg\{ (16\hbar^4 v_F^4 + t^4 |\gamma_{RL}|^4) |\gamma_{RR'}|^2|\gamma_{LR'}|^2 + \\
 &4zt\hbar v_F(t^2|\gamma_{RL}|^2-4\hbar^2 v_F^2)(|\gamma_{LR'}|^2-|\gamma_{RR'}|^2) \sin(\Omega/2) + \\
 &4t^2\hbar^2 v_F^2\big[|\gamma_{RL}|^2(|\gamma_{RR'}|^4+|\gamma_{LR'}|^4)-4z^2\sin^2(\Omega/2)\big] \bigg\}
\end{split}
\label{noise1} 
\end{equation}
to the leading order in $t'$, and it evidently features oscillations via Pancharatnam's geometric phase like the currents in Eq.~(\ref{Rout}). This set-up, thus, exemplifies an elegant non-interferometric platform where geometric phase of Pancharatnam type is arising from one-particle interference (amplitude interferometry), that can be experimentally detected by simple mesoscopic measurements of current or noise.  Hence, this completes addressing  point (a) and (b) raised in the beginning of this Letter by posting a physical situation which not only supports local and controlled production of Pancharatnam's phase but also its manifestation in physical observables like average current and dc current noise. \\

Finally, we note that $t $ and $\vert \gamma_{LR}\vert$ always appear together as a product in the expressions of both current and noise. This is expected as a finite value of either of these implies breaking of the time reversal invariance on the edges. Additionally, this product will continue to a be a single parameter in our set-up as long as the inter-edge bias $V_L-V_R=0$ since it preserves the symmetry between the left and the right moving edge. \\


\noindent
{\underline{\it Pancharatnam phase in intensity interferometry}}: With this backdrop, we now address point (c) mentioned above. We study  a set-up comprising of HES which is simultaneously coupled to two SPE's at the same spatial point on the edges such that it provides a {\it{two-source two-detector}} set-up essential for observing intensity interferometry~\cite{samuelsson2004two}. In this case, the current and noise would feature two-particle quadrilateral Pancharatnam loops unlike the triangular loops in the previous case as discussed below.\\
\begin{figure*}[ht] 
\includegraphics[width=2.0\columnwidth]{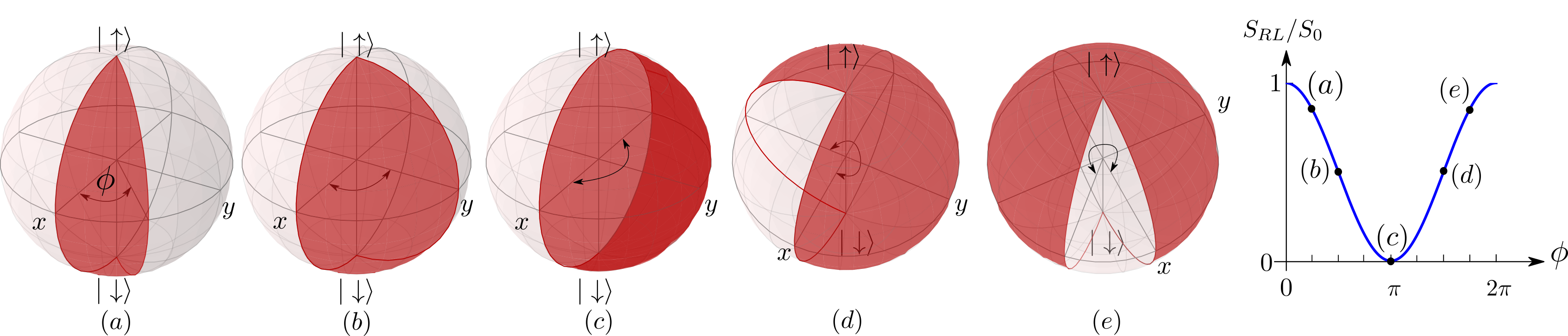}
\caption{(a)-(e) show the evolution of the quadrilateral Pancharatnam loop of Fig.~\ref{fig:figg3} (b) as $\phi$ varies for zero to 2$\pi$. The plot shows the variation of $S_{RL} $ as a function of $\phi$ with $S_0=(-e^3V/h)(t'^4/{\hbar^4 v_F^4})$ being the prefactor in Eq.~(\ref{noise22_simpl}).}
\label{fig:figg4}
\end{figure*}
We start with two SPE's with distinct polarization labeled $R'$ and $L'$  tunnel coupled with the QSHS and their respective spin states are represented by $\vert n_{R'}\rangle$ and $\vert n_{L'}\rangle$. The tunneling Hamiltonian has the same form as Eq.~(\ref{htun}) except that $\eta,\eta'\in\{R,L,R',L'\}$. We further assume the tunneling strength for both the SPE's to be the same ($t'$).  The average currents and the noise are calculated with a voltage bias $V$ applied to both the SPE's while the edge states are kept grounded. The current expressions $\langle I_\eta^{\rm out} \rangle = \frac{e^2V}{h} \sum_{\eta'} |s_{\eta \eta'}|^2$, where $\eta$ can be $R$ or $L$ and $\eta'\in\{R',L'\}$, when explicitly written by substituting the corresponding $S$-matrix elements, take the forms
\begin{equation}
\begin{aligned}
& \langle I_{(L/R)}^{\rm out} \rangle = \frac{e^2V}{h} \bigg\{\frac{t'^2}{\hbar^2 v_{F}^2}\big(|\gamma_{(L/R)R'}|^2+ |\gamma_{(L/R)L'}|^2\big) +\\
& \frac{t'^4}{2\hbar^4 v_{F}^4}\big(|\gamma_{LR'}|^2|\gamma_{RR'}|^2+|\gamma_{LL'}|^2|\gamma_{RL'}|^2 + (|\gamma_{(L/R)L'}|^2+ \\
& |\gamma_{(L/R)R'}|^2)^2 + \gamma_{RR'}\gamma_{R'L}\gamma_{LL'}\gamma_{L'R} + {\rm h.c.}  \big) \bigg\} + \mathcal{O}(t'^6), 
\label{LRout1}
\end{aligned}
\end{equation}
where we have considered time reversal symmetric edge states i.e. $\langle n_L | n_R \rangle =0$. It should be noted that the presence of local back scattering ($t \neq 0$) is of no consequence for current $\langle I_{(L/R)}^{\rm out} \rangle$ as long as  $\langle n_L | n_R \rangle =0$ on the edges and $V_L-V_R=0$ is maintained. Also from Eq.~(\ref{LRout1}), we observe that Pancharatnam loops appear only in $t'^{4}$ order unlike the case of single SPE taking form of geodesic quadrilateral on the Bloch sphere with the four states in the order $R\rightarrow R'\rightarrow L\rightarrow L'\rightarrow R$ [see Fig.~\ref{fig:figg3} (b)]. 
Similarly, the cross-correlated noise between $R$ and $L$ obtained to $t'^4$ order (which is the leading order) reads as
\begin{equation}
\begin{aligned}
 S_{RL} = {} &-\frac{e^3V}{h} \frac{t'^4}{\hbar^4 v_{F}^4} \bigg\{ \vert\gamma_{R^{'}R}\vert^2\vert\gamma_{LR^{'}}\vert^{2} 
 + \vert\gamma_{L^{'}R}\vert^2\vert\gamma_{LL^{'}}\vert^{2}\\
  &+ \gamma_{RR'}\gamma_{R'L}\gamma_{LL'}\gamma_{L'R}+{\rm h.c.}\bigg\}.
\end{aligned}
\label{noise22}
\end{equation}
In this equation, the last term (and its h.c.), which represents a quadrilateral Pancharatnam loop, has a clear interpretation in terms of two-electron interference~\cite{blanter2000shot} where the two-particle amplitude for ``SPE $R'$ shooting an electron at the edge $R$ and SPE $L'$ shooting another electron at the edge $L$ simultaneously" is interfering with the two-particle amplitude for ``SPE $R'$ shooting an  electron at the edge $L$ and SPE $L'$ shooting another electron at the edge $R$ simultaneously". This is precisely the reason why the leading order contribution to cross-correlated noise comes at forth order in $t'$.  \\
Now to observe neat manifestations of Pancharatnam phase in currents and noise we start by considering an explicit choice for the spinors involved;  $\vert n_{R}\rangle = [1 \hspace{0.15cm} 0]^{T}$ and $\vert n_{L}\rangle = [0 \hspace{0.15cm} 1]^{T}$ which can be represented by the north and south pole of the \bloch sphere [see Fig.~\ref{fig:figg3} (b)]. Next we consider one of the SPEs' magnetization to be directed along the x-axis so that $\vert n_{R^{'}}\rangle = [1 \hspace{0.15cm} 1]^{T}/\sqrt{2}$ and the other SPE's magnetization is kept tunable in the $x$-$y$ plane which could gives rise to oscillations in current and noise via the variation of Pancharatnam's geometric phase. We represent its spin state as $\vert n_{L^{'}}\rangle = [1\hspace{0.15cm}e^{\iota\phi}]^{T}/\sqrt{2}$. Then the expressions for the currents and noise reduce to 
\begin{equation}
 \begin{split}
  \langle I_L^{\rm out} \rangle &= \langle I_R^{\rm out} \rangle \\ &= \frac{e^2V}{h} \bigg\{\frac{t'^2}{\hbar^2 v_{F}^2}+\frac{t'^4}{\hbar^4 v_{F}^4}\bigg(1+\cos^2{\frac{\Omega^{'}}{4}}\bigg)\bigg\},
 \end{split}
\label{curr_2_tips_simpl}
\end{equation}
and 
\begin{equation}
\begin{split}
S_{RL}&=-\frac{e^3V}{h}\frac{t'^{4}}{\hbar^4 v_F^4}\cos^2{\frac{\Omega^{'}}{4}},
\end{split}
\label{noise22_simpl}
\end{equation}
where, $\Omega^{'}=2\phi$ is the solid angle subtended by the geodesic quadrilateral formed by the spin states $\vert n_{R}\rangle$, $\vert n_{L}\rangle$, $\vert n_{R'}\rangle$ and $\vert n_{L'}\rangle$ at the center of the \bloch sphere. Hence, by tuning $\phi$, one can induce oscillations in the noise whose origin lies purely in two-particle type Pancharatnam's geometric phase as shown in Fig.~\ref{fig:figg4}. These oscillations have mild effects in current as the leading order contribution  appears in order $t'^{2}$ while the $\Omega^{'}$ dependent terms appear in the sub leading order. On the other hand, in case of cross-correlated noise they have dominant effects as they appear in the leading order itself yielding neat oscillations in noise as a function of $\phi$ as shown in Fig.~\ref{fig:figg4}. \\ 

\begin{figure}
\includegraphics[width=0.7\columnwidth]{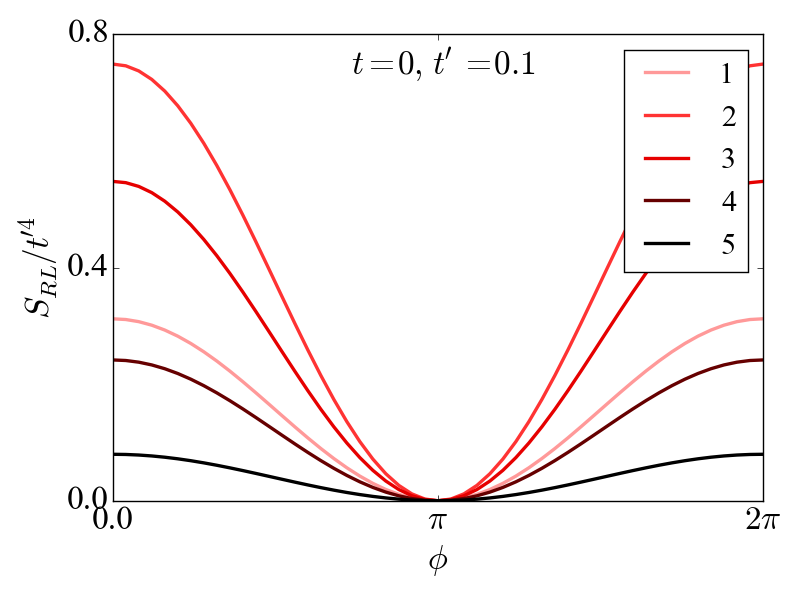}
\caption{The noise [scaled by a factor of $t'^{-4}$ to compare with Eq.~(\ref{noise22_simpl})] in the intensity interferometry set-up measured as a function of the Pancharatnam phase $\phi$ as mentioned in Eq.~(\ref{noise22_simpl}) (with $\phi=\Omega'/2$) by including multiple tunneling points (the number shown in the legend) which reveals that these oscillations indeed survive orbital dephasing.}
\label{fig:figg5}
\end{figure}


\noindent
{\underline{\it Discussion}}: The central finding of our Letter is realization of a two-particle interference pattern arising from Pancharatnam-type geometric phase owing to spin dynamics alone. Further more our result for the amplitude interferometry involving a single SPE is not merely a stepping stone to arrive at the main goal of this work  of obtaining multi-particle interference but is interesting in its own right. Note that for the case of single SPE, the leading order contribution to current [Eq.~(\ref{Rout})] which features Pancharatnam loops is $\mathcal{O}(t')^2$ while inclusion of another SPE pushes it to $\mathcal{O}(t')^4$ [Eq.~(\ref{LRout1})]. For the noise, however, the leading order with Pancharatnam loops remains the same - $\mathcal{O}(t')^4$ [see Eq.~(\ref{noise1}) and Eq.~(\ref{noise22})]. If we assume a situation for the single SPE set-up in which the spin state of the SPE lies in the $x-y$ plane such that it is making an equal angle with that of the spin of right and left moving edge states i.e. $|\gamma_{RR'}| = |\gamma_{LR'}|\equiv\alpha$, one arrives at a neat expression for the physically measurable quantity like current asymmetry defined as $\mathcal{R}\equiv\langle I_R^{\rm out} \rangle-\langle I_L^{\rm out} \rangle=(e^2V/h)\mathcal{R}_0 \sin(\Omega/2)$. Its Pancharatnam's phase dependence is similar to that appearing in the expression for the cross-correlated noise [Eq.~(\ref{noise22_simpl})]. Also, the expression for noise [Eq.~(\ref{noise1})]  for the single SPE simplifies considerably leading to $S_{RL}=(e^3V/h)(S_{RL}^0+S_{RL}^1\cos\Omega)$ (all the constants $\mathcal{R}_0, S_{RL}^0$ and $S_{RL}^1$ are functions of $\alpha,t/\hbar v_F$ and $t'/\hbar v_F$). Hence, the single SPE also stands as an interesting set-up for observing Pancharatnam type geometric phase both in terms of current and noise. \\

As the mechanism to produce the interference pattern is local, it is expected to be robust and immune to the spatial dephasing in the system which we have explicitly verified in the two-SPE set-up corresponding to the intensity interferometry. We include multiple tunneling points (up to 5) into the two-SPE set-up and numerically estimate the noise after performing an averaging over the dynamical phases picked up by the electrons while traversing between consecutive tunneling points, hence, providing a model for an extended junction with an inbuilt orbital dephasing.  The results are presented in Fig.~\ref{fig:figg5}. The plot evinces the robustness of the oscillations against orbital dephasing. This fact can be of great importance as it can serve as a boon while exploring entanglement generation in such a set-up by postselection~\cite{roychowdhury2016quantum} in the context of two-particle interferometers. \\


\noindent
{\underline{\it Conclusion}}: In the present Letter, for the first time, by exploiting the spin dynamics of electrons alone, we propose a scenario involving a solid state setting which allows for purely local generation and manipulation of two-particle interference pattern arising from Pancharatnam-type geometric phase that can be interpreted as multi-particle A-B effect in spin space. The possibility of local generation and manipulation of two-particle interference pattern could provide a completely new route to solid state interferometry. Last but not the least, it is worth noting that the mechanism to produce the interference pattern being purely local could provide an enhanced immunity to bulk sources of spatial dephasing as demonstrated in the Letter.\\


\noindent
{\underline{Acknowledgments}}: SD would like to thank Michael V. Berry for enlightening conversations related to the Pancharatnam phase, N. Mukunda 
for clarifying some subtleties associated with representing multi-photon polarization states on a Poincar\'e sphere~~\cite{chaturvedi2017classical} and discussion on possible implications for electronic counterpart, and  Mikhail Kiselev for useful discussion. 
The work of PM is supported by funding from the University Grants Commission under the second phase of University with Potential of Excellence (UPE II) and DST-PURSE grant from the Department of Science and Technology at JNU. SD would like to acknowledge  Ministry of Human Resource Development for generous funding and IISER Kolkata for the research facilities.
The authors would like to acknowledge the kind hospitality from KITP (KR), Theory Group at CERN (PM),  ICTP under the associate scheme (SD) during crucial stages of this work.

\bibliographystyle{apsrev}
\bibliography{references}

\end{document}